\documentclass[10pt, preprint]{emulateapj}

\shorttitle{Propagating slow magnetoacoustic waves in coronal loops}
\shortauthors{Wang, Ofman, \& Davila}

\begin{document}

\title{Propagating slow magnetoacoustic waves in coronal loops observed by Hinode/EIS}

\author{T. J. Wang and L. Ofman\altaffilmark{1}\\
Department of Physics, Catholic University of America and NASA Goddard Space
Flight Center, Code 671, Greenbelt, MD 20771; wangtj@helio.gsfc.nasa.gov
and\\
J. M. Davila\\
NASA Goddard Space Flight Center, Code 671, Greenbelt, MD 20771}

\altaffiltext{1}{Visiting Associate Professor, Tel Aviv University, Israel}

\begin{abstract}
We present the first Hinode/EIS observations of 5 min quasi-periodic oscillations detected in 
a transition-region line (He~{\small II}) and five coronal lines (Fe~{\small X}, Fe~{\small XII},
Fe~{\small XIII}, Fe~{\small XIV}, and Fe~{\small XV}) 
at the footpoint of a coronal loop. The oscillations exist throughout the whole observation, 
characterized by a series of wave packets with nearly constant period, typically persisting 
for 4-6 cycles with a lifetime of 20-30 min. There is an approximate in-phase relation between 
Doppler shift and intensity oscillations. This provides evidence for slow magnetoacoustic 
waves propagating upwards 
from the transition region into the corona. We find that the oscillations detected in 
the five coronal lines are highly correlated, and the amplitude decreases 
with increasing temperature. The amplitude of Doppler shift oscillations decrease by a factor
of about 3, while that of relative intensity decreases by a factor of about 4 
from Fe~{\small X} to Fe~{\small XV}. These oscillations may be caused by the leakage of the 
photospheric p-modes through the chromosphere and transition region into the corona, 
which has been suggested as the source for intensity oscillations previously observed by TRACE. 
The temperature dependence of the oscillation amplitudes can be explained by damping of the waves 
traveling along the loop with multithread structure near the footpoint. Thus, this property
may have potential value for coronal seismology in diagnostic of temperature structure in a
coronal loop.     
\end{abstract}


\section{Introduction}

A variety of propagating and standing MHD waves (e.g., slow mode, Alfv\'{e}n, and fast mode) have 
been observed in the Sun's outer atmosphere. They are mainly in coronal loops, but also in other 
structures such as coronal plumes and prominences \citep[see reviews by][]{asc04, wan04, wan05, nak05, 
dem05, ban07}. The detection of MHD waves in the solar corona is crucial to determine the
presence and relevance of wave-based heating mechanisms since they are an obvious candidate to
transport energy from the solar surface into the solar atmosphere.
Such observations may also be used to improve existing estimates of coronal properties,
both from direct measurements and indirect methods such as {\it coronal seismology} 
\citep[see, e.g.,][]{rob84, nak99, nak01, rob03, nak05}.

The first detection of propagating slow magnetoacoustic waves was made in Ultraviolet Coronagraph Spectrometer 
(UVCS) on board the Solar and Heliospheric Observatory (SOHO) observations of coronal holes, 
high above the limb by \citet{ofm97, ofm00a}. Similar compressive disturbances, with amplitudes 
of the order of 10$-$20\%, and periods of 10$-$15 min, were detected in polar plumes by 
\citet{def98} with Extreme-ultraviolet Imaging Telescope (EIT)/SOHO. \citet{ofm99, ofm00b} 
identified the observed compressive disturbances as propagating, slow magnetoacoustic waves, 
damped by compressive viscosity.

Similar intensity disturbances propagating along active-region loops with speeds of 
122$\pm$44 km~s$^{-1}$ and intensity variations of 4.1\%$\pm$1.5\% were observed in 
Transition Region and Coronal Explorer (TRACE) Fe~{\small IX/X} 171 \AA\ data 
\citep{nig99, dem00, rob01, dem02a, dem02b, mce06} and EIT/SOHO Fe~{\small XII} 195 \AA\ data \citep{ber99}. 
These disturbances were also interpreted as slow magnetoacoustic waves \citep{nak00}.
\citet{dem02a} found a distinct difference in dominant periods of waves in loops situated above
sunspots (172$\pm$32 s) and those above plage regions (321$\pm$74 s). This difference suggests
that these waves probably originate from the underlying oscillations, i.e., the 3 min
chromospheric/transition-region oscillations in sunspots and the 5 min solar global
oscillations ($p$-modes). 

Many authors have found evidence for the existence of 3 min sunspot oscillations propagating 
through the chromosphere and transition region into the lower corona 
\citep[e.g.][]{bry99a, bry99b, bry02, flu01, osh02, ren03}. 
The amplitude of the oscillations is found to reach a peak in the transition region lines 
then decrease with increasing temperature. These results are confirmed
by a recent study of \citet{mar06} who clearly showed that the 3 min umbral transition region 
oscillations are directly connected to the 3 min wave propagation along the TRACE loops. 

\begin{figure*}
\epsscale{1.0}
\plotone{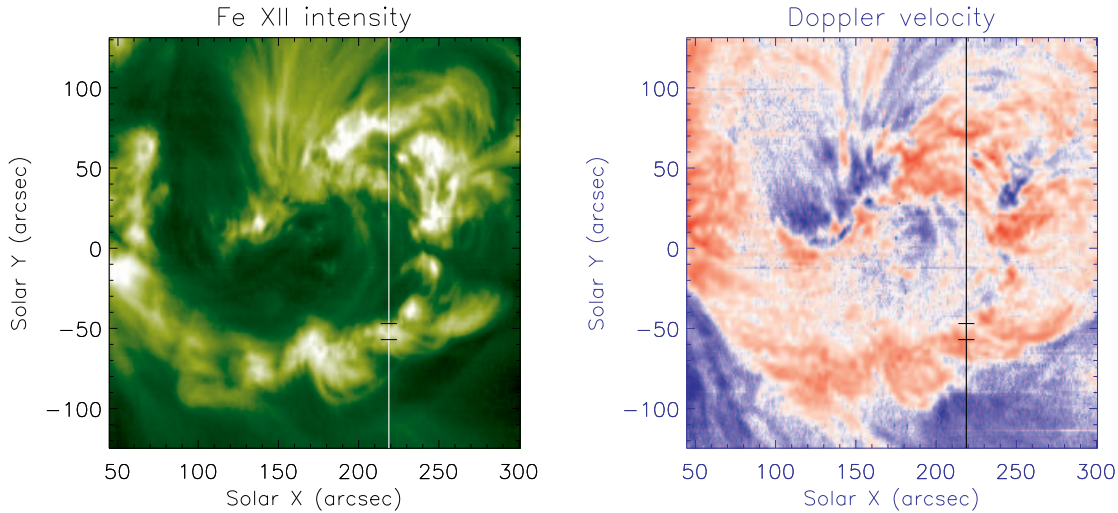}
\caption{ \label{fgmap}
{\it Left panel:} The intensity map in the Fe XII 195.12 \AA\ line from EIS. The raster observation was
taken from 10:42 to 11:52 UT on 2007 Feburary 2. {\it Right panel:} The Doppler velocity measurements. 
The red color represents the red shift and the blue color the blue shift with a scale range from 
$-$20 km s$^{-1}$ to $+$20 km s$^{-1}$. The vertical line in each plot
shows the position of the EIS 1$^{''}$ slit during the sit-and-stare observation, which was taken from
12:39 to 23:12 UT. The short horizontal lines on the slit mark the range of rows showing oscillations. 
The coordinate system for the figure corresponds to the end time of the raster observation. 
}
\end{figure*}

The $p$-mode oscillations are normally evanescent because their periods are well above the 
cutoff period in the upper photosphere and low chromosphere as well as transition region in non-magnetic 
solar atmosphere \citep[e.g.,][]{erd07}. For magnetoacoustic gravity waves, however, \citet{bel77} 
predicted that in regions of low-$\beta$ plasma, the highly inclined magnetic field can channel 
the low-frequency waves from the photosphere into the overlying coronal atmosphere due to the 
increase of the cutoff period. Numerical simulations based on this theory have demonstrated the appearance 
of 5 min waves in chromospheric spicules and in coronal loops near active regions \citep{dep04, dep05}. 
The presence of this effect in the lower chromosphere was confirmed recently by many authors with 
the TRACE 1700 and 1600 \AA\ data \citep{mci06, jef06, blo06, vec07}. In particular, \citet{fon93a} 
and \citet{jef06} pointed out the importance of the observed low-frequency ($<$5 mHz) waves which 
may provide a significant source of energy for heating solar chromosphere. Moreover, the detection 
of these waves propagating from the low atmosphere into the corona is also needed for our 
understanding of the energy balance in the outer solar atmosphere. Unfortunately, such observations
are very few. Only \citet{mar03} report the observation of 5 min propagating oscillations 
simultaneously at chromospheric, transition region and coronal temperatures. 

The EUV Imaging Spectrometer (EIS) onboard {\it Hinode} can simultaneously capture many emission
lines from transition region to coronal temperatures, providing us a good opportunity to study 
the temperature-dependent behavior of the oscillations and the propagation of waves in the solar
atmosphere \citep{mari08} and is highly valuable for coronal seismology \citep{van08, erd08}. 
In this study we report the first Hinode/EIS observation of the 5 min upwardly propagating 
slow magnetoacoustic waves simultaneously at transition region and coronal temperatures in a plage region.
The oscillations show up in both Doppler shift and line intensity with an amplitude dependent
on the temperature. Section 2 describes the observations. Two examples of oscillation packets are
analyzed in Sect. 3 and the properties of their temperature dependence are presented in Sect. 4.
Interpretation and discussion are given in Sect. 5, and finally our conclusions in Sect. 6.

  \begin{deluxetable}{lcc}
 \tablecaption{Emission lines observed with EIS.   \label{tablin}}
 \tablewidth{0pt}
 \tablehead{ 
 \colhead{Ion}    &  \colhead{Wavelength (\AA)} &  \colhead{Log $T_{max}$(K)}}
 \startdata
 He {\small II}   & 256.32 & 4.90 \\
 Fe {\small X}    & 184.54 & 5.98 \\
 Si {\small X}    & 258.37/261.04  & 6.13 \\
 Fe {\small XII}  & 195.12 & 6.13 \\
 Fe {\small XIII} & 202.04 & 6.20 \\
 Fe {\small XIV}  & 264.78/274.20 & 6.28 \\
 Fe {\small XV}   & 284.16 & 6.32 \\
 Fe {\small XVI}  & 262.98 & 6.43 \\
 \enddata
\end{deluxetable}

\section{Observations}

\subsection{Data}
An overall description of EIS is available in \citet{cul07}, and the {\it Hinode} mission is described by 
\citet{kos07}. EIS has both imaging
(40${''}$ and 266${''}$ slots) and spectroscopic (1${''}$ and 2${''}$ slits) capabilities, in the wavelength
range of 170$-$210 \AA\ and 250$-$290 \AA\ with high spectral (0.0223\AA) and spatial resolution 
(1${''}$ pixel$^{-1}$). Its spectroscopic mode can operate in a rastering mode
(repeated exposures while scanning over the observation target) or a sit-and-stare mode (repeated exposures
at the same spatial location).

The observations analyzed in this study cover the central portion of NOAA active region 10940 and were
obtained on 2007 February 2, when it was located close to the disk center. An EIS spectroheliogram was taken
from 10:42 UT to 11:52 UT with the 1${''}$ slit and covering a $256{''}\times256{''}$ region \citep{war08}. 
The exposure times were 15 s. This data set contained 20 spectral windows. 

\begin{figure*}
\epsscale{1.0}
\plotone{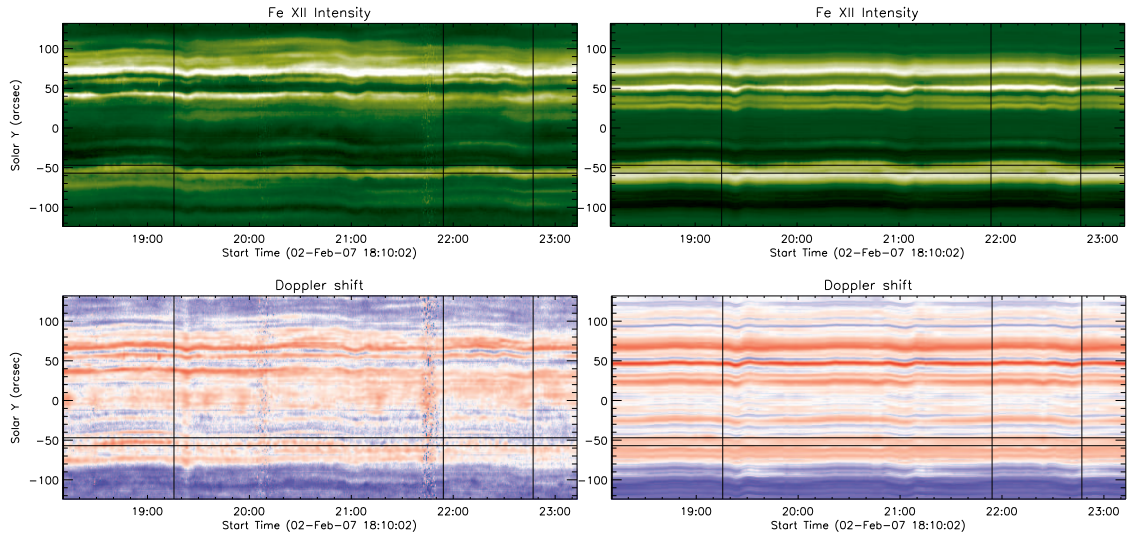}
\caption{ \label{fgss}
{\it Top left:} Time series of the EIS sit-and-stare intensity in the Fe XII 195.12 \AA\ line. 
{\it Bottom left:} The corresponding Doppler shift measurements. {\it Right panels:} The constructed 
time series of the Fe XII intensity and Doppler shift, based on the raster image and spacecraft 
drifts determined from the housekeeping data for XRT. The horizontal lines in each plot mark 
the range of rows showing oscillations. The vertical lines mark two time ranges for oscillations
analysed (i.e., 18:10$-$19:15 UT and 21:54$-$22:47 UT). The Doppler shifts are plotted in a scale range 
from $-$20 km s$^{-1}$ to $+$20 km s$^{-1}$. 
}
\end{figure*}

A sit-and-stare observation within the region began at 12:39 UT and consisted of 1200 exposures with 
the 1${''}$ slit, each with an exposure time of 30 s. Each exposure covered 20 data windows on the EIS
detectors. This paper presents results for the lines in ten of those windows. Each window was 24 spectral 
pixels wide and covered a height of 400${''}$ in the north-south direction. Table~\ref{tablin} lists
the emission lines included in this study and their temperatures of formation.

The raw data were processed by the standard SolarSoft routine {\em eis\_prep} to remove detector bias
and dark current, hot pixels, and cosmic rays, and calibrated using the prelaunch absolute calibration.
The EIS slit tilt and orbital variation in the line centroids were also removed from the data. The
emission lines in each spectral window were then fitted with Gaussian profiles, providing the total
intensity, the Doppler shift, and the line width. The data in the short-wavelength detector were 
shifted downward by 18.5 pixels in the y-direction to correct for the offset between the two detectors.
  
Figure~\ref{fgmap} shows the intensity and Doppler shift maps of the active region covered by the EIS
spectroheliogram in the Fe~{\small XII} line. A string of brightenings in a circular shape are dominated by
the redshifted flow. The location of the EIS slit for the sit-and-stare observation is across
two loop systems. The large one at about $y$=$0{''}-100{''}$ is most clearly seen in the 
Fe~{\small XV} and Fe~{\small XVI} lines \citep[see][]{war08}, indicating that it consists of hot 
loops with the temperature of 2$-$3 MK. The patches of brightening seen in Fe~{\small XII} appear 
to locate at the footpoint region of these hot loops. 

The small one at the region from $y$=$-60{''}$ to $y$=$-50{''}$ is most clearly seen in the 
Fe~{\small XI}$-$Fe~{\small XIV} lines, indicating a temperature of 1$-$2 MK. The sit-and-stare observation,
made at the footpoint of this small loop, reveals oscillations in both Doppler shift and intensity
in many emission lines. We will present the detailed analysis of the oscillation in Sects.~\ref{sctosc} and
\ref{sctmln}. Data from the GOES X-ray monitors show no flare-like fluctuations above the B level during
the EIS observation.

\begin{figure}
\epsscale{1.0}
\plotone{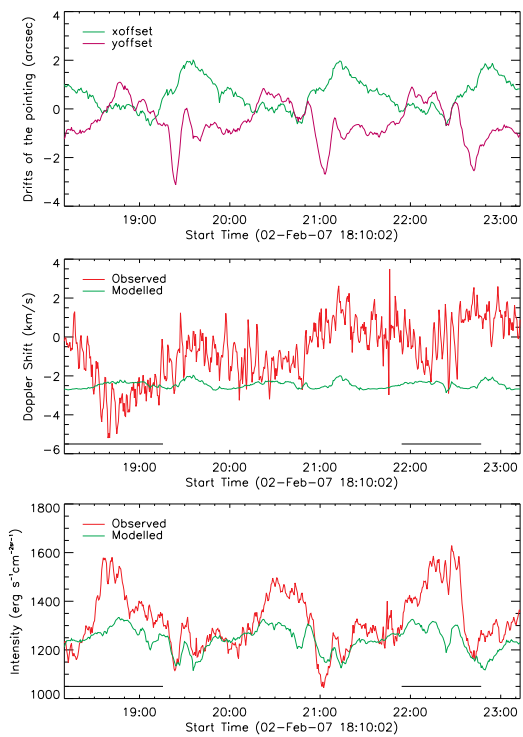}
\caption{ \label{fgjit}
{\it Top panel:} Drifts of the spacecraft pointing in x and y directions, obtained from the 
housekeeping data for XRT. {\it Middle panel:} Average time profiles of the observed and modelled Doppler 
shifts for a strip shown in Fig.~\ref{fgss}. Here positive values for the Doppler shift represent
the blueshifted emission. {\it Bottom panel:} Average time profiles of the observed and modelled intensities.
In the middle and bottom panels the thick horizontal lines mark two periods, over which the
oscillations are analyzed with the wavelet method shown in Figs.~\ref{fgwv1} and \ref{fgwv2}.}
\end{figure}

\begin{figure*}
\epsscale{1.0}
\plotone{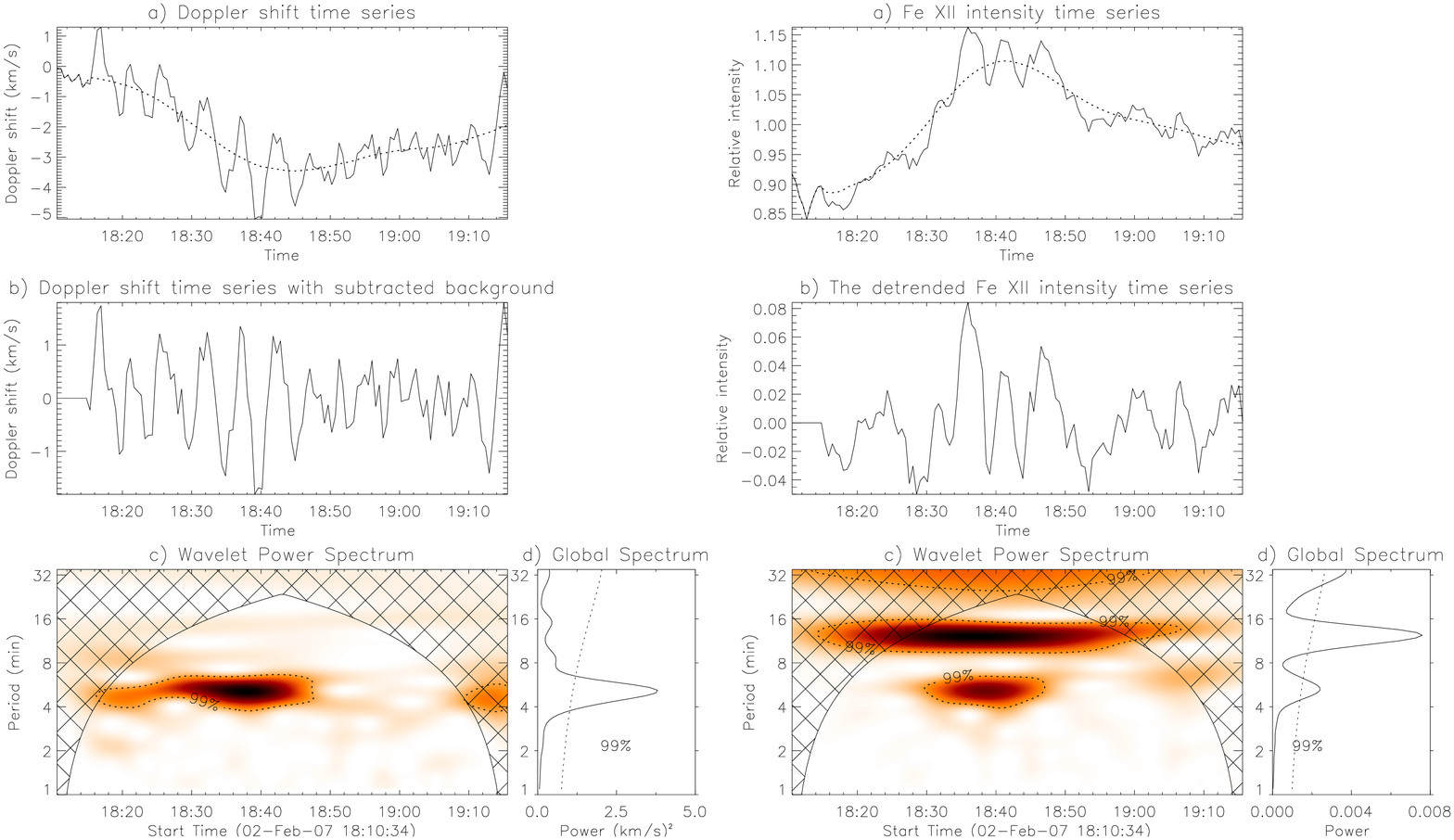}
\caption{ \label{fgwv1}
Wavelet analysis for averaged Doppler shift and intensity time series in the Fe XII 195\AA\ line 
from 18:10 to 19:15 UT. 
{\it Left Panels:} (a) The Doppler shift data (solid line) and the background trend (dotted line). 
(b) The detrended Doppler shift data. (c) The wavelet power spectrum. The dark color represents high 
power and the dotted contour encloses regions of greater than 99\% confidence for a white-noise process. 
Cross-hatched regions on either end indicate the ``cone of influence", where edge effects become important. 
(d) The global wavelet spectrum (solid line) and its 99\% confidence level (dotted line). 
{\it Right Panels:} Same as the left panels but for averaged intensity time series.}
\end{figure*}

\subsection{Effect of instrumental jitter}
Hinode is known to have instrumental jitter in both $x$ and $y$ directions that can have a range of
up to 4${''}$ during the observation. We first examine the jitter effect during the sit-and-stare
observation. The left panels of Figure~\ref{fgss} show time series of the intensity and measured Doppler
shift for the Fe~{\small XII} line along the EIS slit. We found that three evident ``dip''-like features
occurred at about 19:23, 21:02, and 22:40 UT. Since such ``dip''  features are seen simultaneously
at three bright structures at the positions of $y\sim-$60${''}$, 40${''}$, and 70${''}$, we can determine
that they were caused by the jitter of several pixels in the $y$ direction. No high frequency, large amplitude
jitter in the $y$-direction are found in other times. The quasi-periodic Doppler shift oscillation  
was found to be associated with the small loop at $y\sim-$60${''}$. The oscillating
structure has a width along the slit over more than 10${''}$ and the oscillation is visible over the
entire observation. These facts indicate that the detected oscillation could not be caused by jitter 
in the $y$ direction. 

However, if there is a high gradient in brightness and Doppler shift across the slit, jitter in
the $x$ direction may cause artificial oscillation. To examine if this is the case, we model the
sit-and-stare observation based on the EIS spectroheliogram  
and the drift of the XRT pointing which is obtained using the SolarSoft routine {\em xrt\_jitter}. 
The XRT pointing is a good proxy for EIS pointing although not perfect. Since no XRT data are available
before about 18:04 UT, we choose to model and analyze the sit-and-stare data set taken after 18:04 UT.
The top panel of Figure~\ref{fgjit} shows the displacements of the XRT pointing determined from 
the housekeeping data. Both the $x$- and $y$-displacements are mainly orbitally varying, with amplitudes 
up to 2${''}$-3${''}$. The sit-and-stare observation is modelled by extracting the slit slices
from the EIS spectroheliogram taken at 10:42 UT (see Fig.~\ref{fgmap}) with the position of 
the slit varying with time, whose drifts in $x$- and $y$-directions are taken as the displacements 
of the XRT pointing. Since the intensity and Doppler shift distributions are assumed not to change 
with time, the fluctuations shown in the constructed time series (right panels in Fig.~\ref{fgss})
are only caused by the instrumental jitter. The evident ``dip''-like features at 
about 19:23, 21:02, and 22:40 UT which were caused by large displacements in the $y$-direction are 
consistent well with the EIS observation, confirming that the XRT pointing is a good proxy for the EIS.
The orbital variations in brightness at $y$=$-$20${''}$ were also well reproduced. This feature was caused by
the slowly-varying, orbit-related displacements in the $x$-direction which led the EIS slit repeatedly
approaching a small brightening in the west. 

The middle and bottom panels of Figure~\ref{fgjit} show comparisons between the observed and modelled time 
profiles of the intensity and Doppler shift averaged at the loop of interest over 11 pixels from 
$y$=$-$57${''}$ to $-$47${''}$ (see the marked positions in Fig.~\ref{fgss}).
Obviously, the variations in Doppler shift caused by the jitter are too small to account for the observed
oscillation.  Except at the times when the large $y$-displacements occurred, there is
no correlation in intensity variations between the modelled and observed time profiles. Therefore, we can safely
exclude the possibility that the detected oscillation was caused by jitter of the EIS pointing.
In addition, for the EIS sit-and-stare data to be analyzed in the following sections, the pointing drifts 
in the $y$ direction have been removed based on the $y$-displacements observed in XRT.

\section{Analysis of oscillation properties}
\label{sctosc}

The quasi-periodic oscillation in Doppler shift detected at the footpoint of the small loop is clearly seen to exist
during the whole observation (see middle panel of Fig.~\ref{fgjit}). The average root-mean-square (rms) amplitude
of the time series with the subtracted background trend is about 0.7 km~s$^{-1}$. It is clear that this 
oscillation was not related to any flares or impulsive energy release events since the GOES X-ray flux was below 
the B-class level during the observation. We notice that the quasi-periodic time series is characterized by 
a train of oscillations with a nearly constant period on the order of 5 min and an amplitude on the 
order of 1 km~s$^{-1}$, and these oscillations are associated with intensity fluctuations with the same period. 
For example, two typical oscillation periods are 18:10$-$19:15 UT and 21:54$-$22:47 UT. 

\begin{figure*}
\epsscale{1.0}
\plotone{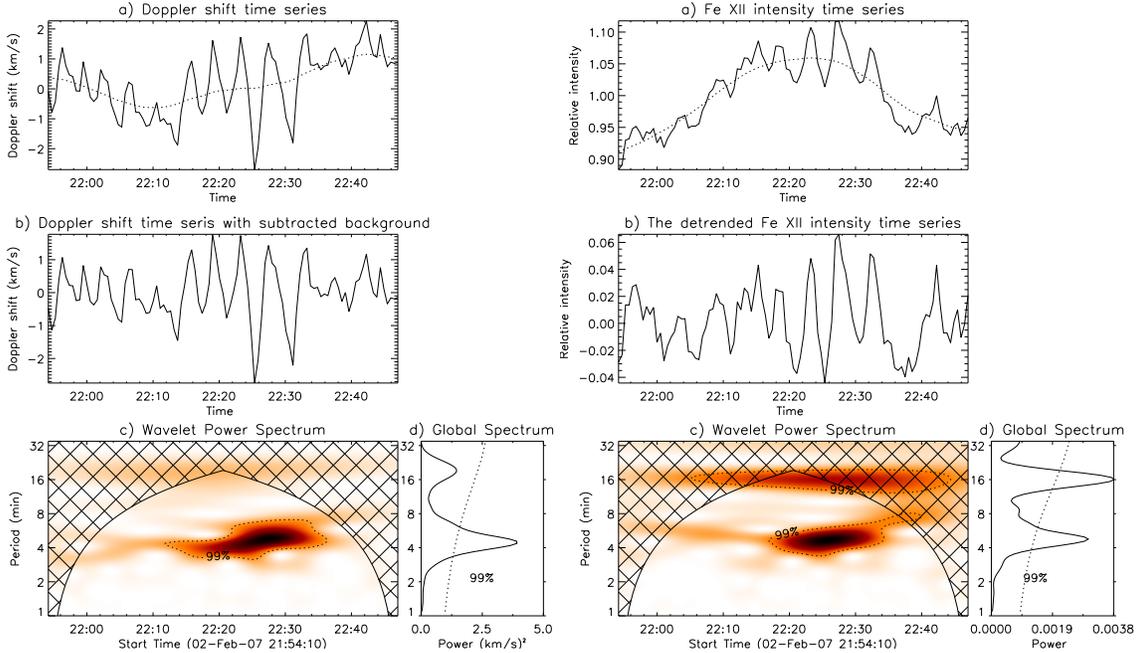}
\caption{ \label{fgwv2}
Wavelet analysis for averaged Doppler shift ({\it left panels}) and intensity ({\it right panels})
time series in the Fe XII 195\AA\ line from 21:54 to 22:47 UT. The annotations are same as 
in Fig.~\ref{fgwv1}.
}
\end{figure*}

In the following we analyze these two oscillations with the wavelet method. The details of the
procedure are given by \citet{tor98}. For the convolution of the time series the Morlet wavelet is chosen,
and to establish whether the oscillations are real, a randomization method is implemented which estimates
the confidence level of the peaks in the wavelet spectrum by assuming the background spectrum as white noise
(with a flat Fourier spectrum). 

\begin{figure}
\epsscale{1.0}
\plotone{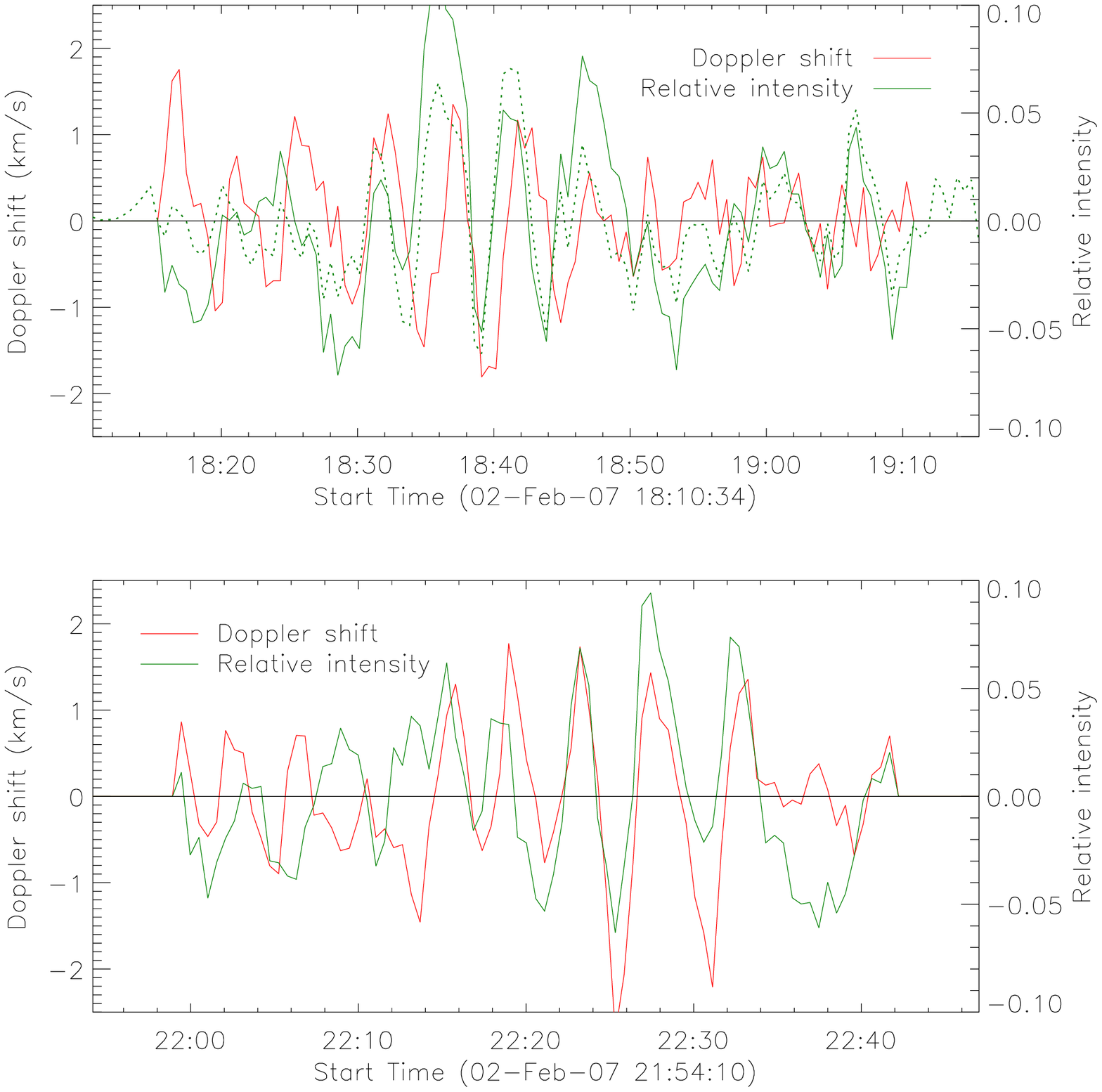}
\caption{ \label{fgphs}
 Phase relationship between Doppler shift and intensity oscillations in the Fe XII line for two wave packets
from 18:10 to 19:15 UT ({\it upper panel}) and from 21:54 to 22:48 UT ({\it bottom panel}). 
In the upper panel, the dotted curve represents the reconstructed intensity time series by subtracting the power 
of the long-period band (8$-$16 min) from the wavelet spectrum. 
Here positive values for the Doppler shift represent the blueshifted emission.
}
\end{figure}

\subsection{Oscillation 1 (18:10$-$19:15 UT)}
\label{subosc1}

The left panel (a) in Figure~\ref{fgwv1} shows the evolution of the Doppler shift in the Fe~{\small XII} 
line averaged over 11 pixels along the slit between $y$=$-$57${''}$ and $-$47${''}$. 
In practice, we first subtract the slowly-varying background trend
from the time series (left panel (b) in Fig.~\ref{fgwv1}). The trend is constructed by using the moving 
average method with a characteristic smoothing time of 10 min. Then the wavelet spectrum and the global 
wavelet spectrum are constructed for the detrended time series (left panels (c) and (d) in Fig.~\ref{fgwv1}). 
The global wavelet spectrum is the average of the wavelet power over time at each oscillation period.
Due to the limited temporal resolution only periods more than 1 min are considered. The wavelet spectrum 
shows strong power at the period in a range of 4$-$6 min over a duration of 6 periods. We measure the
oscillation period as the value where the global wavelet power is peaked and the uncertainty as the half FWHM.
The obtained period, $P_{V1}$, is 5.2$\pm$0.9 min and the amplitude (defined as the square root of the peak global wavelet power), $A_{V1}$, is 1.9 km~s$^{-1}$.

The right panel (a) of Figure~\ref{fgwv1} shows the evolution of the Fe~{\small XII} intensity. The intensity
time series has been normalized by the mean value. A similar wavelet analysis is applied to the detrended
time series (right panel (b) in Fig.~\ref{fgwv1}). The wavelet spectrum shows that most of the power is concentrated within two period bands ranging in 4$-$6 min and 10$-$15 min (right panels (c) and (d) in 
Fig.~\ref{fgwv1}). For the short-period band, we measured the oscillation period, $P_{I1}$=5.2$\pm$1.0 min 
and the relative amplitude, $A_{I1}\approx$5.0\%. The oscillation period is the same as that for the
Doppler shift. The strong power is also seen over a time in agreement with that for the Doppler
shift oscillation. The long-period band has an oscillation period of 12.4$\pm$2.2 min and a
relative amplitude of 8.7\%. No significant power is found at this period in the wavelet spectrum of the 
Doppler shift.

\begin{figure*}
\epsscale{1.0}
\plotone{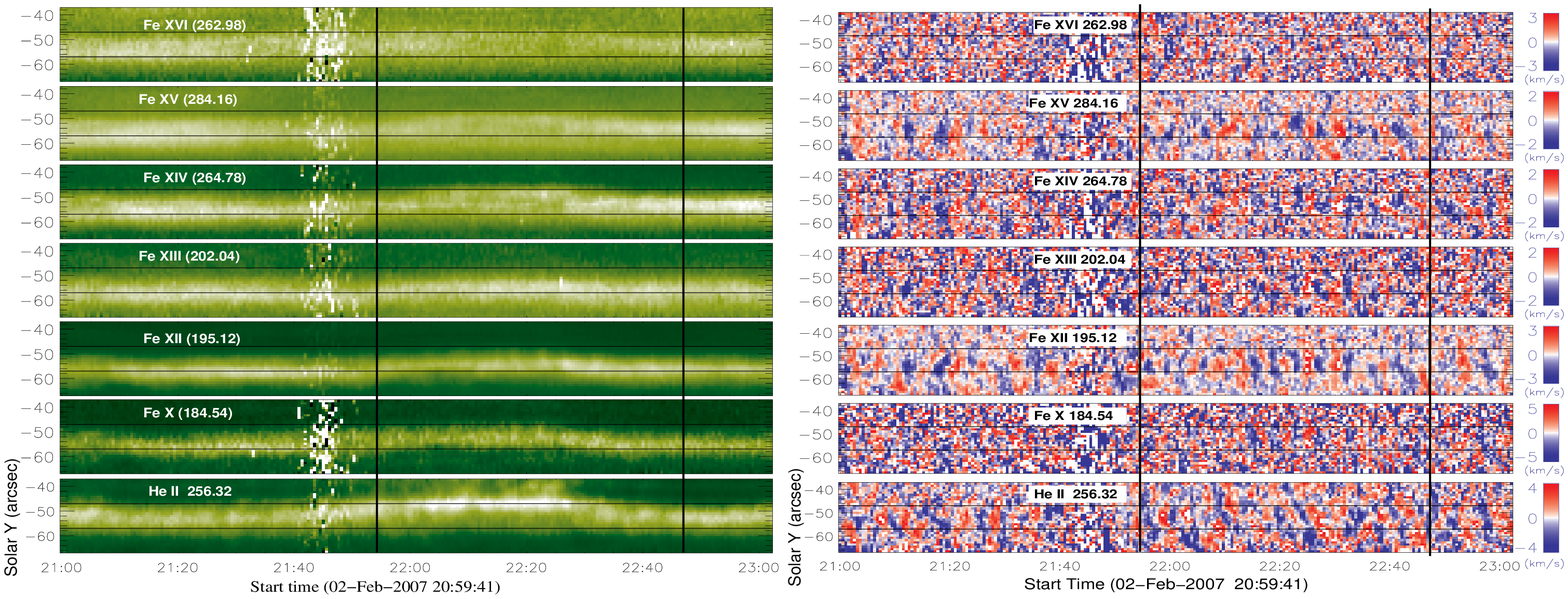}
\caption{ \label{fgssi}
Time series of EIS sit-and-stare intensity data ({\it left panels}) and the Doppler shift measurements
({\it right panels}) in six coronal lines, Fe~{\small X} (184.54 \AA), Fe~{\small XII} (195.12 \AA), 
Fe~{\small XIII} (202.04 \AA), Fe~{\small XIV} (264.78 \AA), Fe~{\small XV} (284.16 \AA), and Fe~{\small XVI}
(262.98 \AA), and a transition region line, He~{\small II} (256.32 \AA). Here the red color represents
the redshift and the blue color the blueshift. The two horizontal lines 
in each panel mark the range of rows showing oscillations. The two vertical lines mark the time range 
for oscillation 2. The pointing drifts in y direction have been corrected in all time series. 
Note that the bad data points seen almost in all lines around 21:45 UT are caused by cosmic rays.
}
\end{figure*}

\subsection{Oscillation 2 (21:54$-$22:47 UT)}
Oscillation 2 also showed up clearly in both the Doppler shift and intensity in the 
Fe~{\small XII} line (Fig.~\ref{fgwv2}). The strong power for the Doppler shift is located at a period range
of 3$-$7 min, with a slight shift from shorter to longer period with time (left panel (c)
in Fig.~\ref{fgwv2}). We measured the oscillation period, $P_{V2}$=4.4$\pm$1.0 min, and the amplitude,
$A_{V2}$=2.0 km~s$^{-1}$, from the global wavelet spectrum using the same method as applied to 
oscillation 1. The wavelet spectrum for the intensity also shows two strong power bands as oscillation 1, 
one covering a shorter period range of 4$-$6 min and the other a longer period range of 13$-$20 min (right panel (c)
in Fig.~\ref{fgwv2}). In contrast to oscillation 1, most of the power for the long-period band of this
oscillation is within the cone of the influence where edge effects become important due to 
the finite length of time series. For the 
short-period oscillation, the period, $P_{I2}$, is 4.8$\pm$0.9 min, and the relative amplitude, 
$A_{I2}$, is 5.4\%. The wavelet analyses for the intensity and the Doppler shift show that the short-period
oscillation occurs in the same frequency range and during the same period. For the long-period oscillation 
in intensity, the period is measured to be 16.1$\pm$3.2 min and the relative amplitude is 6.2\%. 
The measured physical parameters for oscillations 1 and 2 are listed in Table~\ref{tabpar}.

\subsection{Phase Relationship}
\label{sctphs}
We examine the phase relationship between the Doppler shift and intensity oscillations. 
For oscillation 1, we reconstruct
the intensity time series by removing the long-period (12 min) power band from the wavelet spectrum. 
The top panel of Figure~\ref{fgphs} shows that the reconstructed intensity time series (dotted curve) is 
nearly in phase with the original one, and has a phase earlier by about 1/4-period than the Doppler shift 
oscillation. The cross-correlation gives a phase shift of 53$^{\circ}$. 
The bottom panel of Figure~\ref{fgphs} shows a good in-phase relationship between the Doppler shift
and intensity oscillations for oscillations 2. The phase analysis gives the phase shift of 18$^{\circ}$.
Notice that for both oscillations the phase of the intensity oscillation is earlier than the Doppler shift
oscillation. Examination of other cases show that the approximate in-phase relation is more common.

According to linear MHD wave theory, intensity and Doppler shift oscillations are usually associated
with a slow magnetoacoustic longitudinal wave in coronal loops. From the EIS spectroheliogram  the
oscillation is detected near the footpoint of a small loop so that longitudinal motions should have a
line-of-sight component resulting in the observed Doppler shift. The phase relation of oscillation 1
may indicate the presence of a standing wave or two oppositely propagating waves at that time. 
While the approximate in-phase relation for oscillation 2 may indicate that  
the oscillations are dominated by an upwardly propagating wave, but probably overlaid with a 
weak downwardly propagating wave, which caused the small phase shift observed between intensity and
Doppler shift oscillations.

The propagation direction of the wave is determined based on the following linear wave theory. 
Given the axis of $z$ is orientated towards the observer (i.e., the Doppler blueshift [upward motion]
takes positive values), the disturbed velocity of an upwardly propagating linear wave can be 
described in the form,
\begin{equation}
 v(z,t)=V^{\prime}{\rm sin}(kc_st-kz),
 \label{equp}
\end{equation}
while for a downwardly propagating wave the disturbed velocity is,
\begin{equation}
 v(z,t)=V^{\prime}{\rm sin}(kc_st+kz),
 \label{eqdn}
\end{equation}
where $V^{\prime}$ is the amplitude, $k$ the wavenumber and $c_s$ the sound speed. In the above we assumed 
$\omega=kc_s$ with positive $\omega$, $k$, and $c_s$. The linearized continuity
equation is
\begin{equation}
   \frac{\partial\rho^{\prime}}{\partial{t}}+\rho_0\frac{\partial{v}}{\partial{z}}=0,
  \label{eqcn}
\end{equation}
where $\rho_0$ is the background density (a constant), and $\rho^{\prime}$ the density perturbation.
From Eqs.~(\ref{equp}), (\ref{eqdn}) and (\ref{eqcn}) it follows that for an upwardly propagating wave
\begin{equation}
 \rho^{\prime}(z,t)=\left(\frac{V^{\prime}}{c_s}\right)\rho_0{\rm sin}(kc_st-kz), \label{eqdup}   
\end{equation}
and for a downwardly propagating wave
\begin{equation}
 \rho^{\prime}(z,t)=-\left(\frac{V^{\prime}}{c_s}\right)\rho_0{\rm sin}(kc_st+kz). \label{eqddn}
\end{equation}
The above equations indicate that the velocity and density perturbations are in phase for the upwardly
propagating wave, while in opposite phase for the downwardly propagating wave. Clearly, the in-phase relation
for oscillation 2 is consistent with the upwardly propagating wave.

\begin{figure*}
\epsscale{1.0}
\plotone{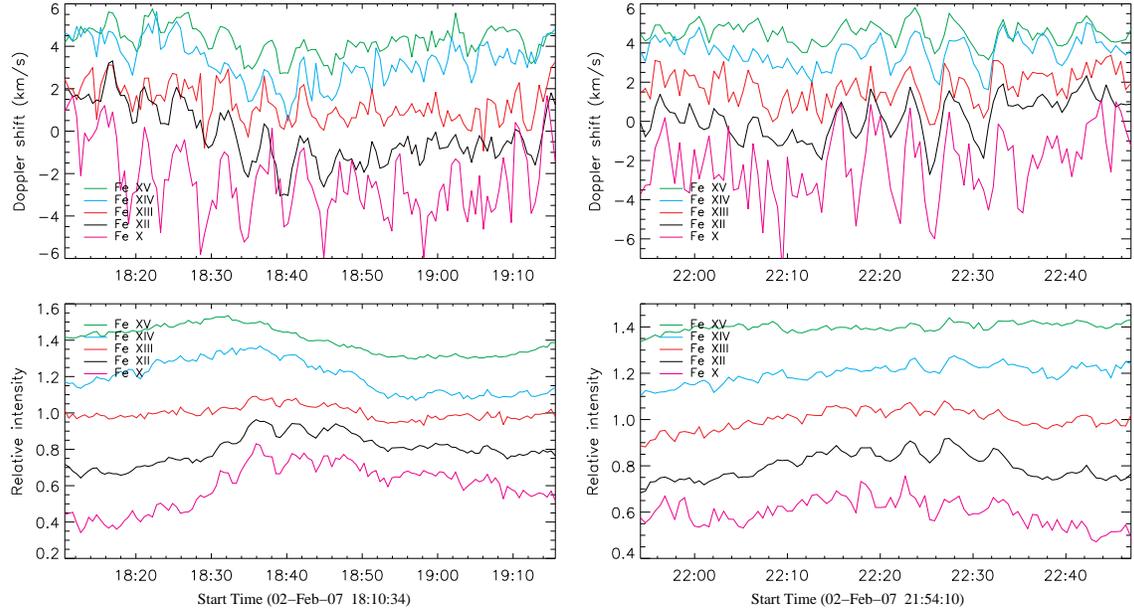}
\caption{ \label{fglc12}
Two selected time series of the averaged Doppler shifts ({\it upper panels}) and intensities ({\it lower panels}) 
in five coronal lines, Fe~{\small X} (184.54 \AA), Fe~{\small XII} (195.12 \AA), Fe~{\small XIII} (202.04 \AA), 
Fe~{\small XIV} (264.78 \AA), and Fe~{\small XV} (284.16 \AA) for a strip marked in Fig.~\ref{fgssi}. 
The zero point of Doppler shifts in each emission line has been arbitrarily shifted in the y-direction
in order to facilitate comparisons. The intensities 
in each emission line have been normalized to the mean averaged over the whole time series, and the light curves
for the Fe~{\small X}, Fe~{\small XII}, Fe~{\small XIII}, Fe~{\small XIV} and Fe~{\small XV} lines are 
shifted by $-$0.4, $-$0.2, 0, 0.2, and 0.4 in the y-direction, respectively.  
}
\end{figure*}

In contrast, the standing acoustic wave shows a quarter-period phase relation, and such examples have been 
observed by the SUMER spectrometer on SOHO \citep{wan03a, wan03b}. In addition, we can exclude the 
possibility that the observed oscillations are caused by a fast sausage-mode wave, because the fast sausage 
mode is characterized by short periods on the order of several$-$tens seconds in coronal loops \citep{asc04}, 
which are not consistent with the dominant period (5 min) of the observed oscillations.  

\begin{figure}
\epsscale{1.0}
\plotone{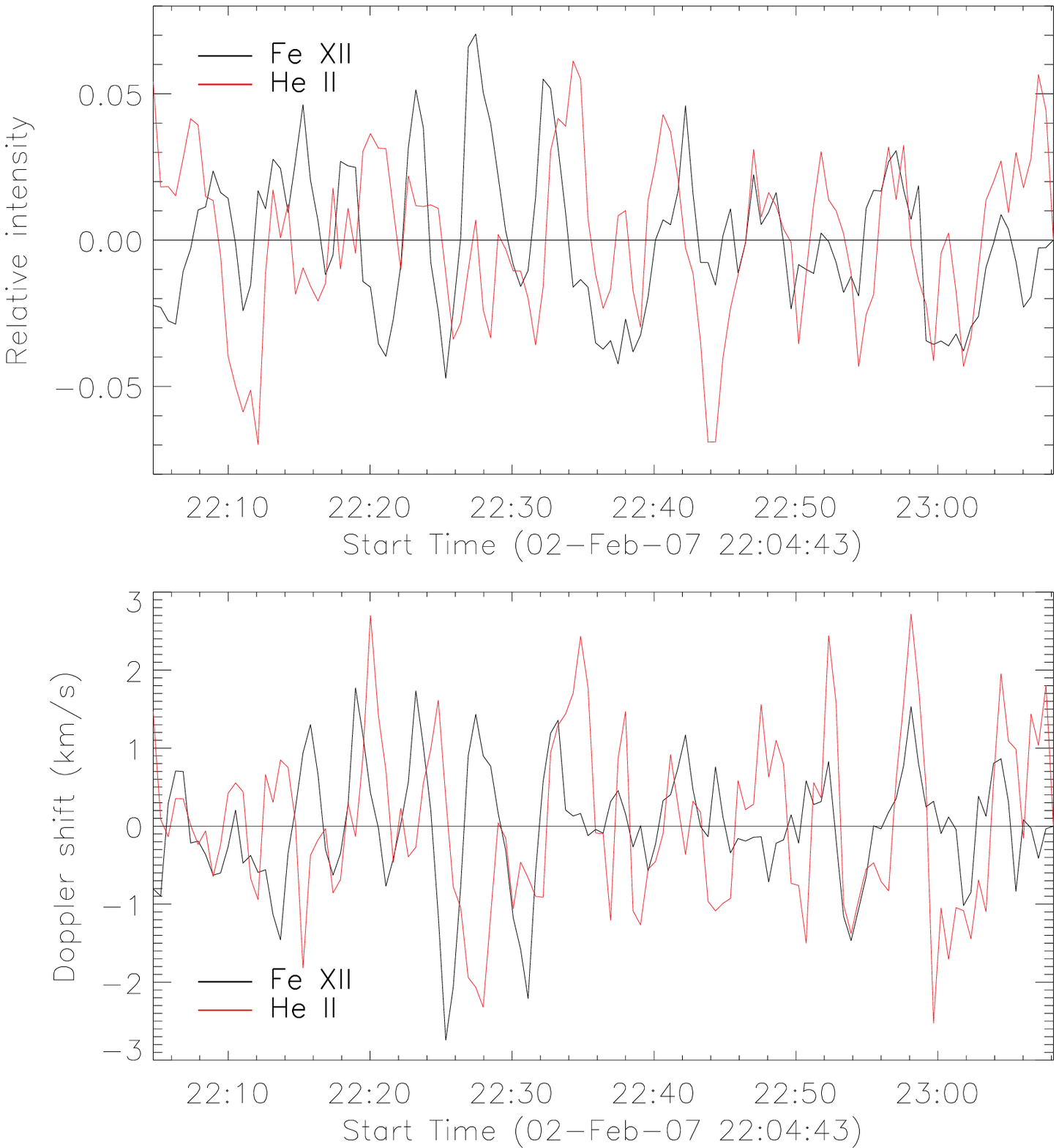}
\caption{ \label{fgfeh}
Comparison between the Fe~{\small XII} (195.12 \AA) and He~{\small II} (256.32 \AA) time series in intensity (upper panel) and Doppler shift (bottom panel) from 22:05 to 23:08 UT.}
\end{figure}

The fast kink mode is nearly incompressible, i.e., with negligible density perturbation, however, \citet{coop03} 
demonstrate that intensity variations can be produced by the kink modes, depending on the viewing angle.
Since the intensity of the emission is proportional to the column depth of the loop, the variation of the LOS
column depth due to the effect of projection causes the variation of the intensity. They find that the observed
amplitude increases with the decreasing wave length and the increasing displacement amplitude of kink perturbations,
and also depends on the angle between the LOS and the axis of the structure. However, the condition in our case 
is not in favor of such an effect because of the very long wavelength (inferred from the period of 5 min), the small 
displacement amplitude (inferred from the Doppler shift amplitude) and the high inclination of the loop.
Provided the LOS is co-planar with the plane of the kink oscillation and has an angle of 45$^{\circ}$ to the axis of 
the loop, we estimate that the intensity amplitude produced by the kink oscillation should be less than 1\% with 
the measured parameters and the theory of \citet{coop03}, which cannot account for the intensity amplitudes observed 
in our study.

\section{Dependence of oscillations on the temperature}
\label{sctmln}

\subsection{Oscillations in Fe~{\small X}$-$Fe~{\small XV}}
We examine the temperature dependence of the oscillation in six coronal emission lines with 
formation temperatures ranging from 1.0 MK to 2.7 MK (see Table~\ref{tablin}).
Figure~\ref{fgssi} shows the evolution of intensity and Doppler shift for the oscillating loop in 
a part of time series. The background trend for Doppler shift time series has been subtracted 
at each position along the slit in order to show the oscillation more clearly. The loop 
is visible in all the six emission lines, but most clearly seen in the Fe~{\small XII}$-$Fe~{\small XIV} lines.
The intensity oscillation is weak and not easily discerned,
while the Doppler shift oscillation can be clearly seen in all lines except for Fe~{\small XVI}.
The quasi-periodic oscillation actually existed over the whole observation. At the time of oscillations
1 and 2, the oscillation appears to be more coherent along the slit (or across the loop) and
more periodic compared to most of the other time.

\begin{figure*}
\epsscale{1.0}
\plotone{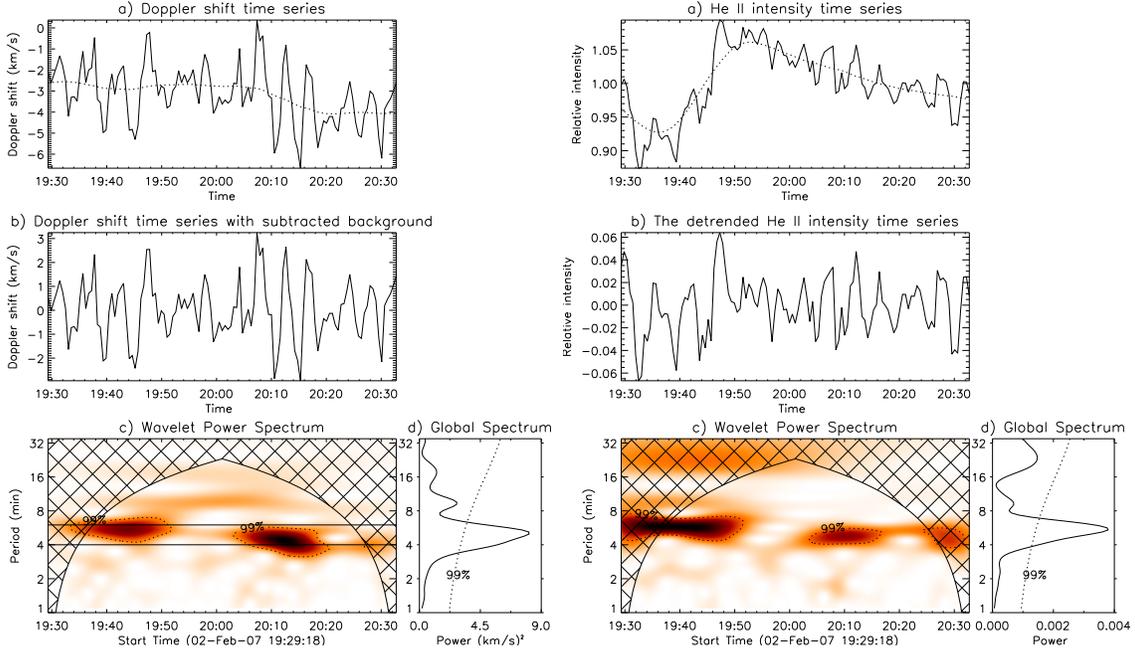}
\caption{ \label{fgwhe}
Wavelet analysis for averaged Doppler shift ({\it left panels}) and intensity ({\it right panels})
time series in the He~{\small II} 256.32 \AA\ line from 19:29 to 20:32 UT. The annotations are same as 
in Fig.~\ref{fgwv1}.
}
\end{figure*}

The top panels of Figure~\ref{fglc12} show comparisons of the averaged Doppler shift profiles over 
11 pixels along the slit between $y$=$-$57${''}$ and $-$47${''}$ in the
Fe~{\small X}$-${\small XV} lines for oscillations 1 and 2. Strikingly, the plots clearly reveal a
dependence of the oscillation amplitude on the plasma temperature, i.e., the amplitude decreases with
increasing temperature. For oscillations 1 and 2, the maximum amplitudes in the Fe~{\small X} line are 
about 2.5 and 3.1 km~s$^{-1}$, while those in the Fe~{\small XV} line are 0.8 and 1.1 km~s$^{-1}$, 
respectively, indicating that the amplitude decreases by a factor of about 3. The oscillations seen at 
different temperatures are nearly in phase. Although the cross correlation shows that the phase of the 
oscillations in Fe~{\small X} is slightly earlier than in Fe~{\small XV}, the shift is measured to 
be within a half exposure time.  The bottom panels
of Figure~\ref{fglc12} show the evolution of the corresponding intensity profiles. For an easier comparison
between the different emission lines, we have normalized the intensity for each line by the mean value
of the time profile and shifted the light curve by a certain value along the $y$ axis. Although the 
intensity oscillation is weak compared to the Doppler shift oscillation, the behavior of the temperature 
dependence of the amplitude is same as for the Doppler shift oscillation. For oscillation 1
three peaks can be seen in the Fe~{\small X} and Fe~{\small XII} lines, but are hardly discerned in
the Fe~{\small XIV} and Fe~{\small XV} lines. For oscillation 2 the decreasing trend of the amplitude
from Fe~{\small X}  to Fe~{\small XV} is most evident for three peaks during 22:22$-$22:34 UT.

 \begin{deluxetable*}{lcccccccccc}
 \tablecaption{Measurements of the oscillations found in the EIS sit-and-stare data in six emission 
 lines\tablenotemark{a}.  \label{tabpar}}
 \tablewidth{0pt}
 \tablehead{ 
 \colhead{Ion}    &  \colhead{$P_{V1}$ } &  \colhead{$A_{V1}$} & \colhead{$P_{I1}$ } &  \colhead{$A_{I1}$} &
  \colhead{$P_{V2}$ } &  \colhead{$A_{V2}$} & \colhead{$P_{I2}$ } &  \colhead{$A_{I2}$} & \colhead{$\sigma$(V)} & \colhead{$\sigma$(I)}\\
   &  \colhead{min} &  \colhead{km/s} & \colhead{min} &  \colhead{\%} & 
   \colhead{min} &  \colhead{km/s} & \colhead{min} &  \colhead{\%} & \colhead{km/s} & \colhead{\%}}
 \startdata
Fe {\small X}    & 5.3$\pm$1.0 & 3.1 & 5.7$\pm$1.4  & 6.8   & 4.3$\pm$0.9 & 3.2 & 4.9$\pm$0.9   & 6.5 & 1.31 & 3.8\\
Fe {\small XII}  & 5.2$\pm$0.9 & 1.9 & 5.2$\pm$1.0  & 5.0   & 4.4$\pm$1.0 & 2.0 & 4.8$\pm$0.9   & 5.4 & 0.72 & 2.2\\
Fe {\small XIII} & 5.1$\pm$1.1 & 1.3 & (4.9$\pm$1.1)& (2.2) & 4.9$\pm$1.2 & 1.4 & (4.3$\pm$1.4) & (2.6) &0.69 & 1.8\\
Fe {\small XIV}  & 5.4$\pm$1.7 & 1.1 & (6.2$\pm$1.4)& (3.1) & 4.6$\pm$1.1 & 1.4 & 4.4$\pm$1.0   & 3.3 & 0.63 & 1.7\\
Fe {\small XV}   & 5.4$\pm$1.1 & 1.2 & (4.5$\pm$0.9)& (1.5) & 4.7$\pm$1.1 & 0.9 & 5.0$\pm$1.1   & 2.0 & 0.48 & 1.1\\
He {\small II}\tablenotemark{b}   & 5.7$\pm$1.0 & 2.4 & 6.1$\pm$1.1 &  6.9   & 4.4$\pm$1.0 & 3.1 & 4.8$\pm$0.9   & 4.8 & 1.07 & 2.7\\
 \enddata
\tablenotetext{a}{ $P_{V1}$ and $P_{I1}$ are periods of the Doppler shift and intensity oscillations
for oscillation 1 (18:10$-$19:15 UT), respectively, while $P_{V2}$ and $P_{I2}$ for oscillation 2 
(21:54$-$22:47 UT). $A_{V1}$ and $A_{I1}$ are amplitudes of the Doppler shift and intensity oscillations
for oscillation 1, respectively, while $A_{V2}$ and $A_{I2}$ for oscillation 2. The value in parentheses 
represents that the peak measured in the global wavelet spectrum is below the 99\% confidence level. 
$\sigma(V)$ and $\sigma(I)$ are the average $rms$ amplitudes for the detrended time series in  
Doppler shift and relative intensity during 18:10$-$ 23:10 UT.}
\tablenotetext{b}{ Note that the two oscillations measured in the He II line are different 
from those measured in Fe~{\small X}$-$Fe~{\small XV}, which correspond to the time series from 19:30 to 19:52 UT 
and from 20:02 to 20:20 UT.}
\end{deluxetable*}

The periods and the amplitudes for the Doppler shift and intensity oscillations in the 
five emission lines are quantitatively measured from the global wavelet spectrum by 
applying the same method as used in Sect.~\ref{subosc1}, and are listed in Table~\ref{tabpar}.
The periods measured in the different lines are nearly same. For oscillation 1 the mean value
of the periods for the Doppler shift in the five lines is 5.2$\pm$0.1 min, and for
oscillation 2 the mean value of the periods is 4.6$\pm$0.2 min. The oscillation amplitude
measured for the Doppler shift from Fe~{\small X} to Fe~{\small XV} decreases by a factor of about 3.
This result obtained with the wavelet method is consistent with that by directly measuring the maximum 
amplitudes mentioned above. For the intensity oscillations the measurement shows that the amplitude 
decreases by a factor of about 4. Note that the intensity oscillations in Fe~{\small XIII}$-$Fe~{\small XV}
for oscillation 1 and that in Fe~{\small XIII} for oscillation 2 are so weak that the peak
measured in the global wavelet spectrum is below the 99\% confidence level. 

\begin{figure}
\epsscale{1.0}
\plotone{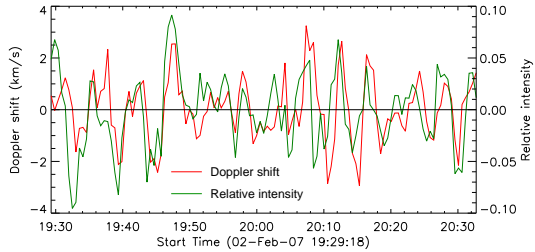}
\caption{ \label{fgphe}
 Phase relationship between Doppler shift and intensity oscillations in the He~{\small II} line for time series
from 19:29 to 20:32 UT ({\it upper panel}). Positive values for the Doppler shift represent the blueshifted emission.
}
\end{figure}

In addition, we can also examine the temperature dependence of the oscillation amplitude by measuring 
the average {\it rms} amplitudes. For the averaged time series of intensity and Doppler shift during 
18:10$-$ 23:10 UT. We first exclude bad data points at about 20:10 UT and 21:45 UT which were caused by cosmic 
rays (see Fig.~\ref{fgssi}), and then subtract the background trend which is taken as the 20-pixel 
smoothing average. Finally, the {\it rms} amplitudes are calculated as the standard deviation for 
the mean of the detrended time series in five coronal lines and are listed in Table~\ref{tabpar}. We 
find that the oscillation amplitudes in Doppler shift and intensity decrease by a factor of 2.7 and 3.5, 
respectively, from Fe~{\small X} to Fe~{\small XV}. This result is in good agreement with that measured 
from oscillations 1 and 2 with the wavelet method.

 \begin{deluxetable}{lccccc}
 \tablecaption{Linear Pearson correlation coefficients between oscillations in Fe~{\small XII} and 
Fe~{\small X} lines and those in Fe~{\small XII} and He~{\small II} lines for intensity ($\rho_I$) 
and Doppler shift ($\rho_V$) over a 1-hour and a 5-hour time series.
  \label{tabcor}}
 \tablewidth{0pt}
 \tablehead{ & \multicolumn{2}{c}{22:05 $-$ 23.08} & &  \multicolumn{2}{c}{18:10 $-$ 23:13}\\
 \cline{2-3} \cline{5-6}\\
 Lines & Fe XII/Fe X & Fe XII/He II & & Fe XII/Fe X & Fe XII/He II}
\startdata
$\rho_I$ & 0.52 & 0.09 & & 0.57 & 0.04\\
$\rho_V$ & 0.61 & 0.12 & & 0.53 & 0.16\\
\enddata
\end{deluxetable}

\subsection{Oscillations in He~{\small II}}
The 5 min quasi-periodic oscillations are also clearly detected in the transition-region line,
He~{\small II} 256.32 \AA, in both intensity and Doppler shift time series over the whole observation
(see bottom panels in Fig.~\ref{fgssi}). Interestingly, we did not find any correlation between the 
oscillations observed in the Fe~{\small X}$-$Fe~{\small XV} lines (e.g.,
oscillations 1 and 2) and those in the He~{\small II} line, even considering the possible time delays.
Figure~\ref{fgfeh} shows comparisons in the detrended relative intensity and Doppler shift between the 
Fe~{\small XII} and He~{\small II} lines. We have removed the contribution of a main blended line, 
Si~{\small X} (256.37 \AA), from the He~{\small II} line intensity by measuring the line intensity 
of Si~{\small X} 261.04 \AA. Indeed, the blended emission from Si~{\small X} can be safely ignored 
for the purpose of the oscillation study because it only contributes about 7\% of the total emission 
of He~{\small II} in the data analyzed. Assuming that the Si~{\small X} line has the same 
relative amplitude in intensity variation as the Fe~{\small XII} line, we estimate that its 
contribution to the intensity variation in He~{\small II} is below 0.4\%. The lack of correlation 
between the oscillations measured in Fe~{\small XII} and He~{\small II} can be clearly seen from 
Fig.~\ref{fgfeh} and the calculated correlation coefficients (see Table~\ref{tabcor}). The 
correlation coefficients between Fe~{\small XII} and He~{\small II} are about 10\% in both intensity 
and Doppler shift, while those between Fe~{\small XII} and Fe~{\small X} are more than 50\% in contrast. 
The similar result is obtained for a 5-hour period of data set. The above analysis 
indicates that the He~{\small II} line detected the oscillation in transition region but not 
in the corona due to the blended emission.     
     
We analyze the oscillations seen in the He~{\small II} line from about 19:30 to 20:30 UT with the 
wavelet analysis. Figure~\ref{fgwhe} shows that two trains of oscillations can be identified in the wavelet 
spectrum and their power distributions in Doppler shift and intensity are similar (panels (c)).
Each oscillation contains about three periods. We measure the oscillation period and amplitude from the 
global wavelet spectrum using the same method mentioned above and list them in Table~\ref{tabpar}. 
We find that the period is in the range 4$-$6 min, and the Doppler shift and relative intensity 
amplitudes are comparable to those measured in the Fe~{\small X} line. Figure~\ref{fgphe} shows 
that the Doppler shift and intensity oscillations are approximately in phase.  Their phase shift measured
with the cross correlation is 28$^{\circ}$. Interestingly, the phase of the intensity oscillation is 
also slightly earlier than that of the Doppler shift oscillation as found in Fe~{\small XII}. 
The approximate in-phase relation between intensity and Doppler shift oscillations indicates the presence of
an upward-propagating slow magnetoacoustic waves in transition region. The small phase shift may 
imply that it is overlaid with a weak downwardly propagating wave as suggested for oscillations
seen in the coronal lines in Sect.~\ref{sctphs}. 

 \begin{deluxetable}{lccccc}
 \tablecaption{Estimates of the wave energy flux for two oscillations measured in six emission 
lines\tablenotemark{a}.  \label{tabflx}}
 \tablewidth{0pt}
 \tablehead{\colhead{Ion} & c$_s$ & V$_1^{'}$ & F$_1$ &  V$_2^{'}$ & F$_2$ \\
         &  km/s &  km/s & 10$^3$ erg/s/cm$^2$ & km/s & 10$^3$ erg/s/cm$^2$}
\startdata
Fe X    & 149  & 5.1 & 7.6$\pm$1.5   & 4.8  & 6.9$\pm$1.4 \\
Fe XII  & 173  & 4.3 & 6.4$\pm$1.3   & 4.7  & 7.5$\pm$1.5 \\
Fe XIII & 191  & 2.1 & 1.7$\pm$0.3   & 2.5  & 2.4$\pm$0.5 \\
Fe XIV  & 210  & 3.3 & 4.4$\pm$0.9   & 3.5  & 5.0$\pm$1.0 \\
Fe XV   & 220  & 1.6 & 1.2$\pm$0.2   & 2.2  & 2.1$\pm$0.4 \\
He {\small II}\tablenotemark{b}   & 43   & $\geqslant$2.4  & $\geqslant$6.2 &  $\geqslant$3.1  & $\geqslant$10. \\
\enddata
\tablenotetext{a}{$c_s$ is the sound speed corresponding to the line formation temperature.  
$V_1^{'}$ and $V_2^{'}$ are the perturbed velocity amplitudes derived from the measured 
relative intensity amplitudes for oscillations 1 and 2 (see Table 2). $F_1$ and $F_2$ are the 
wave energy flux calculated with Eq.~(\ref{eqflx}).}
\tablenotetext{b} {Note that for the He~{\small II} line the low limit of the perturbed velocity 
amplitudes is directly taken from the measured Doppler shift amplitude.} 
\end{deluxetable}

\section{Discussions}
In this paper we have reported the simultaneous detections of 5 min quasi-periodic oscillations in the 
transition-region and coronal lines by Hinode/EIS in a corona loop rooted at plage. 
The oscillations show up in both Doppler shift 
and intensity throughout the whole observation, characterized by a series of trains with a more or 
less constant period lasting for 3$-$6 periods. These oscillation trains are more evident
in Doppler shift than in intensity. The oscillations detected in five coronal lines in the range
Fe~{\small X}$-$Fe~{\small XV} show a high correlation and exhibit a temperature dependence of
the amplitude in both Doppler shift and intensity. The wavelet analyses show that the oscillation power 
is concentrated at the period in a range of 4$-$6 min. Both measurements with the wavelet analysis and
the average {\it rms} method show that the oscillation amplitude decreases by a factor
of about 3 in Doppler shift while decreasing by a factor of about 4 in intensity from Fe~{\small X} to Fe~{\small XV}.
The oscillations measured in the transition region line, He~{\small II}, also have the dominant power 
at the period band of 4$-$6 min, and the amplitudes are comparable to those measured in the Fe~{\small X} line.
No correlation between the oscillations in Fe~{\small X}$-$Fe~{\small XV} and He~{\small II} is found.
The cross correlation shows that the phase of intensity oscillation is slightly earlier (by about 
20$^{\circ}-30^{\circ}$) than the Doppler shift oscillation in Fe~{\small X}$-$Fe~{\small XV} and 
He~{\small II}. The approximate in-phase relation indicates the presence 
of upwardly propagating slow magnetoacoustic waves in both the transition region and the corona
near the footpoint of a loop.

\subsection{The source of waves}
From Eq.~(\ref{eqdup}) we obtain $\rho^{\prime}/\rho_0=V^{\prime}/c_s$.  We examine  
this relation for oscillations 1 and 2 in Fe~{\small XII} with the measurement obtained by the wavelet analysis,
which gives the Doppler shift amplitude, $V^{\prime}_{\parallel}\approx$2.0 km~s$^{-1}$ and the 
relative intensity variation, $I^{\prime}/I_0\approx$5.2\% (see Table~\ref{tabpar}), 
where $I_0$ is the background intensity of the loop. Considering $I^{\prime}/I_0\sim2\rho^{\prime}/\rho_0$ 
and taking $c_s=170$ km~s$^{-1}$ for the formation temperature of Fe~{\small XII}, 
we deduce the expected value of the velocity amplitude, $V^{\prime}\approx$4.5 km~s$^{-1}$. The result
of $V^{\prime}_{\parallel}=0.43V^{\prime}$ is consistent with the fact that the coronal loop, along which the 
slow-mode waves propagate, is highly inclined as seen in Fe~\small{XII} (see Fig.~\ref{fgmap}a).
Assuming that $V^{\prime}_{\parallel}$ is the LOS component of $V^{\prime}$, we estimate that the inclination angle 
of the magnetic field to the vertical is about 65$^{\circ}$. By applying the measured average {\it rms} 
amplitudes, almost the same value for the inclination angle is obtained. The high inclination of coronal fields in
the loop may explain the absence of correlation between the oscillations detected in 
Fe~{\small X}$-$Fe~{\small XV} and He~{\small II} lines supposing the waves travel  
from the transition region to the corona along the loop, because the waves detected in the coronal lines and 
the transition region line should come from the different sources due to the inclination of the coronal fields.

The oscillations we present here show many properties which are very similar to those of upward-propagating 
waves observed in coronal loops associated with plages using the TRACE 171 and 195 \AA\ bandpasses 
\citep[e.g.,][]{dem02a, dem02b, kin03, mar03}. First, these oscillations both have the dominant period of about
5 min and the amplitude of intensity variations of about 4\%$-$5\% (in the Fe~{\small X} and Fe~{\small XII} 
lines for EIS). Second, they are both long-term existing and clearly nonflare-excited. Third, they both appear 
in the footpoints of the highly inclined coronal structures and are detected simultaneously at transition region 
and coronal temperatures. These similarities indicate that the source of these waves is same, i.e., 
the leakage of the $p$-modes through the chromosphere and transition region into the corona. 
In our case the oscillations typically containing wave trains of 4$-$6 cycles with no apparent temporal 
damping are also consistent with the property of the photospheric $p$-modes.
\citet{dep04} have shown that the inclination of magnetic flux tubes can dramatically increase tunneling and
may even lead to direct propagation of the $p$-modes along inclined field lines.

\subsection{Interpretation for the temperature dependence}
Using the EIS data, we have detected the upwardly propagating slow magnetoacoustic waves simultaneously 
in five coronal lines with the formation temperature in the range $\sim$1$-$2 MK. The waves in different 
lines show highly correlation  (almost in phase). Particularly, the temperature dependence of the oscillation 
amplitude is revealed for the first time. The feature that the amplitude decreases with the increasing 
temperature is in good agreement with the observations of 3 min sunspot oscillations which show that the
amplitude peaks in the transition-region lines then decreases with increasing temperature \citep{flu01,
bry02, osh02, mar06}. The local maximum amplitude of waves at the transition region can be explained by
strong stratification driven growth \citep{erd07}, while the decreasing of the amplitude at increasing
temperatures may be relevant to dissipation of slow magnetoacoustic waves in the corona. The propagating 
disturbances observed in the TRACE EUV images were found to be damped very quickly and typically only 
detected in the first 3$-$23 Mm along the loop \citep{dem02b}. Numerical simulations of slow MHD waves 
by \citet{dem03, dem04} showed that a combination of thermal conduction and area divergence yielded
detection lengths that are in good agreement with observed values. \citet{kli04} further developed the model
considering a non-isothermal loop, and found again that thermal conduction plays an important role
in quickly damping of the waves. Here three possibilities are proposed to account for the temperature
dependence of the amplitude for the observed propagating waves.

(1) Assuming that the slow mode waves propagate upwardly in a non-isothermal coronal loop, then the 
smaller amplitude detected at higher temperature lines is because the
lower temperature line forms in a height lower than the higher temperature line and the waves undergo a 
quick damping during the propagation from the low level to the high level. With this 
interpretation time delays are expected to exist between the oscillations detected in the lines of different 
temperatures. Cross-correlation analyses for oscillations 1 and 2 show that the phase of Doppler 
shift time series for Fe~{\small XV} is slightly later than that for Fe~{\small X}, but the
time delay is less than 1 exposure (i.e., $<$30 s). By using the results of the sunspot model of
\citet{obr88}, we estimate the height difference of $\sim$6400 km in vertical direction for the
Fe~{\small X} and Fe~{\small XV} lines. This value is on the same order as estimated from observed 
heights of EUV lines and the coronal model for the active Sun \citep{sim72, sim74}. Taking the sound 
speed of 200 km~s$^{-1}$ and an inclination angle of $\sim$60$^{\circ}$ of the magnetic field to 
the vertical, we estimate the expected time delay to be about 60 s, which is inconsistent with the observed.
On the other hand, the inclination of $\sim$60$^{\circ}$ implies a projected distance of more than 15${''}$ 
on the disk for the part of the loop with a vertical height of 6400 km. Since the slit appears to sit across
the loop (see Fig.~\ref{fgmap}a), this means that the emissions detected with the 1${''}$ slit in Fe~{\small X}
and Fe~{\small XV} lines should not come from the same loop. Thus, this scenario should predict 
no correlation between oscillations detected in the lines of different temperatures (at different heights), 
which contradicts to the observation. In the following we propose the other two alternative scenarios 
based on the loop with multiple threads of different temperatures.

(2) With simultaneous TRACE observations in 171 and 195 \AA\ bandpasses, \citet{kin03} have revealed that 
the slow mode waves propagate outwards along coronal loops of multiple-temperature structure. They
show that the correlation between time series of disturbances observed in the different bandpasses has
a tendency to decrease with distance along the structure. They suggest that the initially high correlation 
may infer the same mechanism for generation of the disturbances observed at different temperatures, while
the decreasing correlation along the loop may be due to phase mixing of the waves. We may explain the
temperature dependence of the oscillation amplitude with a similar picture, since the footpoint of the loop
analyzed in this study is clearly not isothermal as revealed by the EM loci curves for 
Fe~{\small X}$-$Fe~{\small XVI} (not shown). Provided the waves have the 
same amplitudes at the base of the corona and propagate along parallel threads of different temperatures,
we expect the wave of smaller amplitudes detected in higher temperature lines because the dissipation of
slow MHD waves is higher at the hotter plasmas by thermal conduction and compressive 
viscosity, which have been interpreted as the dominate damping mechanism in coronal loops by many 
theoretical studies \citep[e.g.][]{nak00, ofm02, dem04, kli04}. The interpretation in this picture, 
however, still needs to explain the high correlation between the oscillations detected in the
different coronal lines, because time delays are expected if the waves travel at different speeds 
in different threads of the loop having different temperatures. One explanation could be that 
the time delay is too short to resolve
with the 30 s cadence. For example, if assuming that the waves travel over a distance of 
7000 km (about half size of the footpoint brightening) in two threads with the temperature equal 
to the formation temperatures of Fe~{\small X} and Fe~{\small XV}, respectively, the expected time delay 
is only 15 s, less than the exposure time. 

(3) The third picture is proposed specifically to interpret the high correlation between the oscillations 
observed in coronal lines of different formation temperatures supposing that the absence of time delays 
is not due to the limited temporal resolution of the observation. We assume that the propagating waves
are only present in an isothermal coronal loop with the plasma temperature of $\sim$1 MK, which is overlaid
by the hotter ($>$1 MK) loops with no waves propagating inside. The overlying relatively hotter loops 
are shown in the spectroheliogram in the Fe~{\small XV} line (not shown). Then the high correlation
may be explained by the effect of the line response function (emissivity as a function of the temperature)
since the modulated emissions of the different lines come from the same plasma disturbed by the waves. For the
lines with higher formation temperatures, the emission contributed from the 1 MK oscillatory plasma 
becomes less while those from the relatively hotter non-oscillatory plasmas become more 
dominant, therefore, the amplitude of the detected oscillations in both intensity and Doppler shift tend 
to decrease with increasing temperature of the emission lines. However, since both the Fe~{\small X} 
and Fe~{\small XV} lines form in a narrow temperature range, more exactly, the contribution of the emission to 
the Fe~{\small XV} line from a plasma at 1 MK is more than 3 orders of magnitude lower than its peak
emission, the presence of waves only in the 1 MK plasma hardly explains the oscillation amplitude of 1\%$-$2\%
measured for Fe~{\small XV}. In addition, this picture is lack of the theoretical basis for the assumption that
the slow mode waves are only allowed to propagate in the 1 MK cool loops.
 
Therefore, the second scenario provides the best interpretation of the temperature dependence of the 
oscillation amplitude based on the present observations. Higher cadence ($<$ 15 s) observations are suggested
in the future to check the reliability and accuracy of time delays between the oscillations detected 
in coronal lines of different temperatures, which will help us confirm this scenario.  
 
Based on linear wave theory we can explain why the amplitude of relative intensity decreases faster than 
the amplitude of Doppler shift with increasing temperature for the oscillations we observed. From the relation 
of $I^{\prime}/2I_0\approx{V}^{\prime}/c_s$ it follows that 
\begin{equation}
\frac{I^{\prime}/I_0}{V^{\prime}}\propto{T}^{-1/2}, \label{eqivt}
\end{equation}
where $T$ is the plasma temperature of the loop. If assuming $T_1$=1 MK and $T_2$=2 MK, respectively, as 
the temperature of the Fe~{\small X} and Fe~{\small XV} lines, we estimate the 
following ratio theoretically
\begin{equation} 
R_{theo}=\frac{(I^{\prime}/I_0)_{T_1}/(I^{\prime}/I_0)_{T_2}}{V^{\prime}_{T_1}/V^{\prime}_{T_2}}=\left(\frac{T_1}{T_2}\right)^{-1/2}\approx1.4. \label{eqthe}
\end{equation}
This estimate is independent on scenarios proposed for explaining the temperature dependence of the oscillation 
since Eq.~(\ref{eqthe}) is derived from 
the continuity equation. If assuming that the inclination angle of the field line for oscillations detected 
at $T_1$ and at $T_2$ is same, we estimate this ratio
\begin{equation}
R_{obs}=\frac{(I^{\prime}/I_0)_{T_1}/(I^{\prime}/I_0)_{T_2}}{V^{\prime}_{T_1}/V^{\prime}_{T_2}}
       =\frac{(I^{\prime}/I_0)_{T_1}/(I^{\prime}/I_0)_{T_2}}{(V^{\prime}_{\parallel})_{T_1}/(V^{\prime}_{\parallel})_{T_2}}\approx1.3, \label{eqobs} 
\end{equation}
with the observational measurements of $(I^{\prime}/I_0)_{T_1}/(I^{\prime}/I_0)_{T_2}\approx4$ and 
$(V^{\prime}_{\parallel})_{T_1}/(V^{\prime}_{\parallel})_{T_2}\approx3$. A good agreement between $R_{obs}$ and 
$R_{theo}$ provides further support to our interpretation of the observed 5 min oscillations in terms of a slow magnetoacoustic wave. 

For the oscillation in He~{\small II}, we find that the amplitudes of relative intensity and Doppler
shift measured with both the wavelet method and the average {\it rms} method are inconsistent with the
relation of $I^{\prime}/2I_0\sim\rho^{\prime}/\rho_0=V^{\prime}/c_s$. For example, taking $c_s$=43 km~s$^{-1}$ 
we derive the perturbed velocities are 1.5 and 1.0 km~s$^{-1}$ for oscillations 1 and 2, respectively, 
from the measured relative intensities (see Table~\ref{tabpar}), which are evidently smaller than the 
measured line-of-sight velocity amplitudes (2.4 and 3.1 km~s$^{-1}$). 
This inconsistency may imply that the measurements of relative
intensity amplitudes are underestimated, which could be relevant to the complexity of the He~{\small II}
line formation \citep[e.g.][]{fon93b}. Instead, we may estimate the low limit of the true amplitude for
relative intensity from the measurements of Doppler shift amplitudes. 
From $I^{\prime}/I_0\geqslant{2V}^{\prime}_{\parallel}/c_s$, we obtain $I^{\prime}/I_0\geqslant$11\% and 14\%
for oscillations 1 and 2, respectively. The estimated amplitudes are in good agreement with that measured 
by \citet{mar03} for the 5 min oscillation at the footpoint of fan-like TRACE loops. They find amplitudes of
12.4$\pm$2.1\% for the transition region line, O~{\small V}, observed with SOHO/CDS. \citet{bry99a} 
present SOHO/SUMER observations above a sunspot region and find intensity amplitudes of 11\% 
and Doppler velocity amplitudes of 2.7 km~s$^{-1}$ for O~{\small V}. Therefore, our measurements of Doppler
shift amplitudes and the derived intensity amplitudes for He~{\small II} are consistent with the
amplitudes for O~{\small V} found by other studies. Thus, our observations show that the relative intensity 
amplitude of 5 min oscillations decreases from the transition region to the corona. This agrees with
the observations of \citet{flu01, bry02, osh02, mar06} who find that the oscillation amplitude above the
umbra reaches a maximum for emission lines formed close to (1$-$2)$\times10^5$ K, and decreases for higher
temperatures.

\subsection{Estimates of wave energy flux}
We can estimate the energy flux for the propagating waves measured in Fe~{\small X}$-$Fe~{\small XV}
and He~{\small II} lines in the coronal loop by,
\begin{equation}
 F=\frac{1}{2}\rho_0(V^{\prime})^2c_s,
 \label{eqflx}
\end{equation}
where the sound speed is taken as $c_s=152T^{1/2}$ km~s$^{-1}$ with the plasma temperature, $T$, in units of MK,
and $V^{\prime}=(1/2)c_s(I^{\prime}/I_0)$. For the oscillations observed in Fe~{\small X}$-$Fe~{\small XV}, 
we estimate $V^{\prime}$ from the relative intensity amplitudes measured with the wavelet method, while 
for the oscillations in He~{\small II}, we directly take the measured Doppler shift amplitudes as the low 
limit of $V^{\prime}$ for the reason discussed above. We use Si~{\small X} $\lambda$258.37/$\lambda$261.04 
and Fe~{\small XIV} $\lambda$264.79/$\lambda$274.20 density sensitive ratios to diagnose the loop density, 
where the line ratio data are calculated with SSW/CHIANTI v5.2.1. For the data observed from 18:10 to 23:10 UT, 
we obtain the mean electron density as (2.0$\pm$0.4)$\times10^9$ cm$^{-3}$ with the Si~{\small X} 
line ratio, and (1.7$\pm$0.2)$\times10^9$ cm$^{-3}$ with the Fe~{\small XIV} line ratio. We find 
that the values estimated from Si~{\small X} and Fe~{\small XIV} agree well. Note that in measurements of 
the Fe~{\small XIV} $\lambda$274.20 intensity, the emission of a blended line Si~{\small VII} $\lambda$274.18 
has been removed by considering the density-insensitive line ratio, $\lambda$274.18/$\lambda$275.35. 
We find that the blended emission from the Si~{\small VII} $\lambda$274.18 indeed can be ignored in this 
study because it only contributes about 3\% to the Fe~{\small XIV} $\lambda$274.20.
Using $\rho_0$=$1.2m_pN_e$=(4$\pm$0.8)$\times10^{-15}$ g~cm$^{-3}$ (where $N_e$ is taken as the electron
number density measured with Si~{\small X} line ratio, $m_p$ the mass of proton, and a constant of
1.2 due to consideration of coronal He abundance), we estimate the energy flux for 
oscillations 1 and 2 in five coronal lines and list the values in Table~\ref{tabflx}. In estimates of 
the wave energy flux for He~{\small II}, a typical value of mass density for the transition region, 
$\rho_0$=5$\times10^{-14}$ g~cm$^{-3}$, is taken. We find that the energy flux of waves for He~{\small II}
and Fe~{\small X} is on the order of 10$^4$ erg~s$^{-1}$cm$^{-2}$, which decreases to the order of 
10$^3$ erg~s$^{-1}$cm$^{-2}$ for Fe~{\small XV}. Since the coronal radiative energy losses
are typically on an order of $10^6-10^7$ erg~s$^{-1}$cm$^{-2}$ for active regions \citep{asc04}, 
the energy carried by the observed waves is too small to heat coronal loops.

\section{Conclusion}
In conclusion, the upwardly propagating slow magnetoacoustic waves with periods of about 5 min have been 
detected in the transition region and coronal emission lines by Hinode/EIS at the footpoint of a coronal 
loop rooted at plage. The amplitude of the oscillations decreases with increasing temperatures. 
The temperature dependence of the amplitude observed in coronal lines can be explained by the waves
traveling along a loop with multi-thermal temperature structure near the footpoint, and thus this feature
may be valuable for coronal seismology to diagnose the property of multithreads in the loop. 
Many similarities between the waves observed by EIS in this study and the waves observed by TRACE in 
large fan-like loops suggest that the source of the waves is the same, i.e., a leakage of p-modes 
through the temperature-minimum region into the chromosphere and transition region reaching the corona. 
Although the energy carried by these waves is not enough to heat the corona, they may be important
for the heating of the chromosphere, which have been found to be the source for generation of
the periodic spicules in active regions \citep{dep04}.

\acknowledgments
 {\it Hinode} is a Japanese mission developed, launched, and operated by ISAS/JAXA in partnership with
NAOJ, NASA, and STFC (UK). Additional operation support is provided by ESA and NSC (Norway).
TJW is grateful to Drs. Ignacio Ugarte-Utta and David Williams for their valuable discussions and 
suggestions. The authors also thank Dr. Harry Warren and the EIS team in NRL for providing EIS data analysis
tutorials. The work of LO and TJW was supported by NRL grant N00173-06-1-G033. LO was also supported 
by NASA grant NNG06GI55G.

\clearpage


\clearpage
\begin{thebibliography}{}
\bibitem[Aschwanden (2004)]{asc04}
    Aschwanden, M. J. 2004, {\it Physics of the Solar Corona - An Introduction}, Chichester UK: Praxis Publishing Ltd and Berlin: Springer, p. 283
\bibitem[Banerjee et al. (2007)]{ban07} Banerjee, D., Erd\'{e}lyi, R., Oliver, R., \& O'Shea, E., 2007,
       \solphys, 246, 3
\bibitem[Bel \& Leroy (1977)]{bel77} Bel, N., \& Leroy, B. 1977, \aap, 55, 239
\bibitem[Berghmans \& Clette (1999)]{ber99} Berghmans, D. \& Clette, F.
     1999, \solphys, 186, 207
\bibitem[Bloomfield et al. (2006)]{blo06} Bloomfield, D. S., McAteer, R. T. J., Mathioudakis, M., \& 
     Keenan, F. P. 2006, \apj, 652, 812
\bibitem[Brynildsen et al. (1999a)]{bry99a} Brynildsen, N., Leifsen, T., Kjeldseth-Moe, O., et al.
     1999a, \apjl, 511, L121
\bibitem[Brynildsen et al. (1999b)]{bry99b} Brynildsen, N., Kjeldseth-Moe, O., Maltby, P., \& Wilhelm, K.
     1999b, \apjl, 517, L159
\bibitem[Brynildse et al. (2002)]{bry02} Brynildsen, N., Maltby, P., Fredvik, T., \& Kjeldseth-Moe, O.
      2002, \solphys, 207, 259
\bibitem[Cooper et al. (2003)]{coop03} 
    Cooper, F. C., Nakariakov, V. M., \& Tsiklauri, D. 2003, \aap, 397, 765
\bibitem[Culhane et al. (2007)]{cul07}
    Culhane, J. L., et al. 2007, \solphys, 243, 19
\bibitem[DeForest \& Gurman (1998)]{def98} DeForest, C. E., \& Gurman, J. B.,
   1998, \apjl, 501, L217
\bibitem[De Moortel et al. (2000)]{dem00} De Moortel, I., Ireland, J., \& Walsh,
     R. W. 2000, \aap, 355, L23
\bibitem[De Moortel et al. (2002a)]{dem02a} De Moortel, I., Ireland, J.,
    Hood, A. W., \& Walsh, R. W. 2002a, \aap, 387, L13
\bibitem[De Moortel et al. (2002b)]{dem02b} De Moortel, I., Ireland, J.,
    Walsh, R. W., \& Hood, A. W. 2002b,  \solphys, 209, 61
\bibitem[De Moortel \& Hood(2003)]{dem03} De Moortel, I., \& Hood, A. W. 2003, \aap, 408, 755
\bibitem[De Moortel \& Hood(2004)]{dem04} De Moortel, I., \& Hood, A. W. 2004, \aap, 415, 705
\bibitem[De Moortel (2005)]{dem05} De Moortel, I. 2005, Phil. Trans. R. Soc. A, 363, 2743
\bibitem[Erd\'{e}lyi et al. (2007)]{erd07}  Erd\'{e}lyi, R., Malins, C., T\'{o}th, G., \& de Pontieu, B.
    2007, \aap, 467, 1299
\bibitem[De Pontieu et al. (2004)]{dep04} De Pontieu, B., Erd\'{e}lyi, R., \& James, S. P. 2004, Nature, 430, 536
\bibitem[De Pontieu et al. (2005)]{dep05} De Pontieu, B., Erd\'{e}lyi, R., De Moortel, I. 2005, \apjl, 624, L61
\bibitem[Erd\'{e}lyi \& Taroyan (2008)]{erd08} Erd\'{e}lyi, R., \& Taroyan, Y. 2008, \aap, 489, L49
\bibitem[Fludra (2001)]{flu01} Fludra,A. 2001, \aap, 368, 639
\bibitem[Fontenla et al. (1993a)]{fon93a}  Fontenla, J. M., Rabin, D., Hathaway, D. H., \&  Moore, R. L. 
    1993a, \apj, 405, 787
\bibitem[Fontenla et al. (1993b)]{fon93b} Fontenla, J. M., Avrett, E. H., \& Loeser, R. 1993b, \apj, 406, 319
\bibitem[Jefferies et al. (2006)]{jef06} Jefferies, S. M., McIntosh, S. W., Armstrong, J. D.,  et al. 2006, 
     \apjl, 648, L151
\bibitem[King et al. (2003)]{kin03} King, D. B., Nakariakov, V. M., Deluca, E. E., et al. 2003, \aap, 404, L1
\bibitem[Klimchuk et al. (2004)]{kli04} Klimchuk, J. A., Tanner, S. E. M., \& De Moortel, I. 2004, \apj, 616, 1232
\bibitem[Kosugi et al. (2007)]{kos07}
    Kosugi, T., et al., 2007, \solphys, 243, 3
\bibitem[Mariska et al. (2008)]{mari08}Mariska, J. T., Warren, H. P., Williams, D. R., \& 
     Watanabe, T. 2008, \apjl, 681, L41 
\bibitem[Marsh et al. (2003)]{mar03} Marsh, M. S., Walsh, R. W., De Moortel, I., \& Ireland, J.
     2003, \aap, 404, L37
\bibitem[Marsh \& Walsh (2006)]{mar06} Marsh, M. S., \&  Walsh, R. W. 2006, \apj, 643, 540
\bibitem[McEwan \& de Moortel (2006)]{mce06} McEwan, M. P., \& de Moortel, I. 2006, \aap, 448, 763
\bibitem[McIntosh \& Jefferies (2006)]{mci06} McIntosh, S. W., \& Jefferies, S. M. 2006, \apjl, 647, L77
\bibitem[Nakariakov et al. (1999)]{nak99}
    Nakariakov, V. M., Ofman, L., DeLuca, E. E., et al. 1999,  Science, 285, 862
\bibitem[Nakariakov \& Ofman (2001)]{nak01} Nakariakov, V. M., \& Ofman, L.
   2001, \aap, 372, L53
\bibitem[Nakariakov et al. (2000)]{nak00} Nakariakov, V. M., Verwichte, E.,
    Berghmans, D., \& Robbrecht, E. 2000, \aap, 362, 1151
\bibitem[Nakariakov \& Verwichte (2005)]{nak05}
   Nakariakov, V. M., \& Verwichte, E. 2005, {\it Living Reviews in Solar Physics},  2, 3,
   (http://www.livingreviews.org/lrsp-2005-3)
\bibitem[Nightingale et al. (1999)]{nig99} Nightingale, R. W., Aschwanden, M. J., \& Hurlburt, N. E.
    1999,\solphys, 190, 249
\bibitem[Obridko \& Staude (1988)]{obr88} Obridko, V. N., \& Staude, J., 1988, \aap, 189, 232
\bibitem[Ofman et al. (1997)]{ofm97} Ofman, L., Romoli, M., Poletto, G.,
   et al. 1997, \apjl, 491, L111
\bibitem[Ofman et al. (1999)]{ofm99} Ofman, L., Nakariakov, V. M., \&
   Deforest, C. E. 1999, \apj, 514, 441
\bibitem[Ofman et al. (2000a)]{ofm00a} Ofman, L., Romoli, M., Poletto, G.,
   et al. 2000a, \apj, 529, 592
\bibitem[Ofman et al. (2000b)]{ofm00b} Ofman, L., Nakariakov, V. M., \&
    Sehgal, N. 2000b, \apj, 533, 1071
\bibitem[Ofman \& Wang (2002)]{ofm02} Ofman, L., \& Wang, T. J. 2002, \apjl,  580, L85
\bibitem[O'Shea et al. (2002)]{osh02} O'Shea, E., Muglach, K., \& Fleck, B. 2002, \aap, 387, 642
\bibitem[Rendtel et al. (2003)]{ren03} Rendtel, J., Staude, J., \& Curdt, W. 2003, \aap 410, 315
\bibitem[Robbrecht et al. (2001)]{rob01} Robbrecht, E., Verwichte, E.,
       Hochedez, J. F., et al. 2001, \aap, 370, 591
\bibitem[Roberts et al. (1984)]{rob84}
   Roberts, B., Edwin, P. M., \& Benz, A. O. 1984, \apj, 279, 857
\bibitem[Roberts \& Nakariakov (2003)]{rob03}
   Roberts, B., \& Nakariakov, V. M. 2003 {\it NATO Advanced Workshop: Turbulence, Waves and Instabilities}
   eds. R. Erd\'{e}lyi et al., Kluwer Academic Publishers, p. 167
\bibitem[Simon \& Noyes (1972)]{sim72} Simon, G. W., \& Noyes, R. W. 1972, \solphys, 22, 450
\bibitem[Simon et al. (1974)]{sim74} Simon, G. W., Seagraves, P. H., Tousey, R., et al. 1974, \solphys, 39, 121
\bibitem[Torrence \& Compo (1998)]{tor98} Torrence, C., \& Compo, G. P. 1998, Bull. Meteor. Soc., 79, 61
\bibitem[van Doorsselaere et al. (2008)]{van08} van Doorsselaere, T., Nakariakov, V. M., Young, P. R., \&
    Verwichte, E. 2008, \aap, 487, L17
\bibitem[Vecchio et al. (2007)]{vec07} Vecchio, A., Cauzzi, G., Reardon, K. P., et al. 2007,
   \aap, 461, L1
\bibitem[Wang et al. (2003a)]{wan03a}
    Wang, T. J., Solanki, S. K., Innes, D. E., et al. 2003a, \aap, 402, L17
\bibitem[Wang et al. (2003b)]{wan03b}
    Wang, T. J., Solanki, S. K., Curdt, W., et al. 2003b, \aap, 406, 1105
\bibitem[Wang (2004)]{wan04}
   Wang, T. J. 2004, in {\it Proc. of SOHO 13, Waves, Oscillations and Small-Scale Transient Events in the Solar
   Atmosphere: A Joint View from SOHO and TRACE}, ed. H. Lacoste, ESA SP-547, 417
\bibitem[Wang (2005)]{wan05}
   Wang, T. J. 2005, in {\it Proc. of the International Scientific Conference on Chromospheric and Coronal Magnetic
   Fields}, eds. D. E. Innes, A. Lagg, S. K. Solanki et al., ESA SP-596, 42
\bibitem[Warren et al. (2008)]{war08} Warren, H. P., Winebarger, A. R., Mariska, J. T., et al. 2008,
  \apj, 677, 1395

\end{thebibliography}
\end{document}